\documentclass[sigconf,nonacm,anonymous=false]{acmart}

\definecolor{MAGENTA}{rgb}{1.0, 0.0, 1.0} 
\definecolor{darkgreen}{rgb}{0.0, 0.7, 0.2} 
\definecolor{darkyellow}{rgb}{0.8, 0.7, 0.3} 
\usepackage{textcomp}%
\usepackage{booktabs}%
\usepackage{amsmath}%
\usepackage{rotating}%

\usepackage{tabu}%
\usepackage{multirow}%

\newcommand{\lowcount}[1]{\textcolor{lightgray}{#1}}

\newcommand{\fourprompts}[0]{Multi-Criteria}
\newcommand{\llama}[0]{LLaMA}
\newcommand{\exact}[0]{Exactness}

\newcommand{\cover}[0]{Coverage}
\newcommand{\Cover}[0]{Coverage}
\newcommand{\topic}[0]{Topicality}
\newcommand{\Topic}[0]{Topicality}
\newcommand{\fit}[0]{Contextual Fit}
\newcommand{\Fit}[0]{Contextual Fit}

\newcommand{\llmevaluation}[0]{LLM-evaluation}

\newcommand{\llmjudge}[0]{LLMJudge}
\newcommand{\dltwenty}[0]{DL20}
\newcommand{\dlnineteen}[0]{DL19}
\newcommand{\llamasmall}[0]{\llama{}-3-8B}
\newcommand{\llamabig}[0]{\llama{}-3.3-70B}
\newcommand{\tfive}[0]{FLAN-T5-large}

\usepackage{placeins} 
\usepackage{enumitem}
\setlist[description]{leftmargin=1em}

\begin{document}

\title{Criteria-Based LLM Relevance Judgments}

\author{Naghmeh Farzi}
\email{Naghmeh.Farzi@unh.edu}

\affiliation{%
   \institution{University of New Hampshire}
   \city{Durham}
   \state{NH}
   \country{USA}
}
\author{Laura Dietz}
\email{dietz@cs.unh.edu}

\affiliation{%
   \institution{University of New Hampshire}
   \city{Durham}
   \state{NH}
   \country{USA}
}

\begin{abstract}
 Relevance judgments are crucial for evaluating information retrie\-val systems, but traditional human-annotated labels are time-con\-suming and expensive. As a result, many researchers turn to automatic alternatives to accelerate method development. Among these, Large Language Models (LLMs) provide a scalable solution by generating relevance labels directly through prompting. However, prompting an LLM for a relevance label without constraints often results in not only incorrect predictions but also outputs that are difficult for humans to interpret.

We propose the \fourprompts{} framework for LLM-based relevance judgments, decomposing the notion of relevance into multiple criteria---such as exactness, coverage, topicality, and contextual fit---to improve the robustness and interpretability of retrieval evaluations compared to direct grading methods. We validate this approach on three datasets: the TREC Deep Learning tracks from 2019 and 2020, as well as LLMJudge (based on TREC DL 2023). Our results demonstrate that \fourprompts{} judgments enhance the system ranking/leaderboard performance.
Moreover, we highlight the strengths and limitations of this appro\-ach relative to direct grading approaches, offering insights that can guide the development of future automatic evaluation frameworks in information retrieval.\footnote{\textbf{Online Appendix at 
% \url{https://anonymous.4open.science/r/criteria-judge}
\url{https://github.com/TREMA-UNH/appendix-criteria-based-llm-relevance-judgments/}
}}
\end{abstract}


\maketitle

\section{Introduction} 

The evaluation of information retrieval (IR) systems relies on relevance judgments to assess the quality of retrieved passages. By determining which pieces of information are relevant to specific information needs (queries), we derive evaluation scores for different IR systems. These scores allow us to compare systems and select the most effective approaches. For a benchmark of queries and a corpus, the evaluation process depends on having relevance labels for each query and document (or passage) pair in the corpus.

While human-assigned relevance labels remain the gold standard, their acquisition at scale presents significant challenges in terms of cost, time, and consistency. Ultimately, only human judges can be the arbitrator of relevance \cite{Faggioli2023PerspectivesOL,soboroff2024dontusellmsmake,clarke_llm-based_2024}, but we aim to provide methods to support the rapid development of new information retrieval approaches.

\bigskip

\noindent We study the task of automatically predicting passage-level relevance labels that can be used with standard IR evaluation tools.

\medskip
\textbf{Task Statement.} Given a query and a passage, predict the relevance on a graded scale.
The objective is to produce relevance labels that, when used to evaluate a set of retrieval systems, yield a system ranking (i.e., leaderboard) similar to the one derived from human-annotated judgments.

\medskip
\textbf{Example Domain.} While our formulation is task-agnostic, all experiments are conducted on the TREC Deep Learning (DL) benchmark.
The TREC DL uses a 0--3 relevance scale, where 3 indicates perfect relevance and 0 denotes no relevance.
Retrieval systems submitted to this benchmark are ranked using gold-standard labels.
Our goal is to predict relevance labels such that the resulting leaderboard closely mirrors that of the human judgments.

\begin{figure}
    \centering
    \includegraphics[width=1\linewidth]{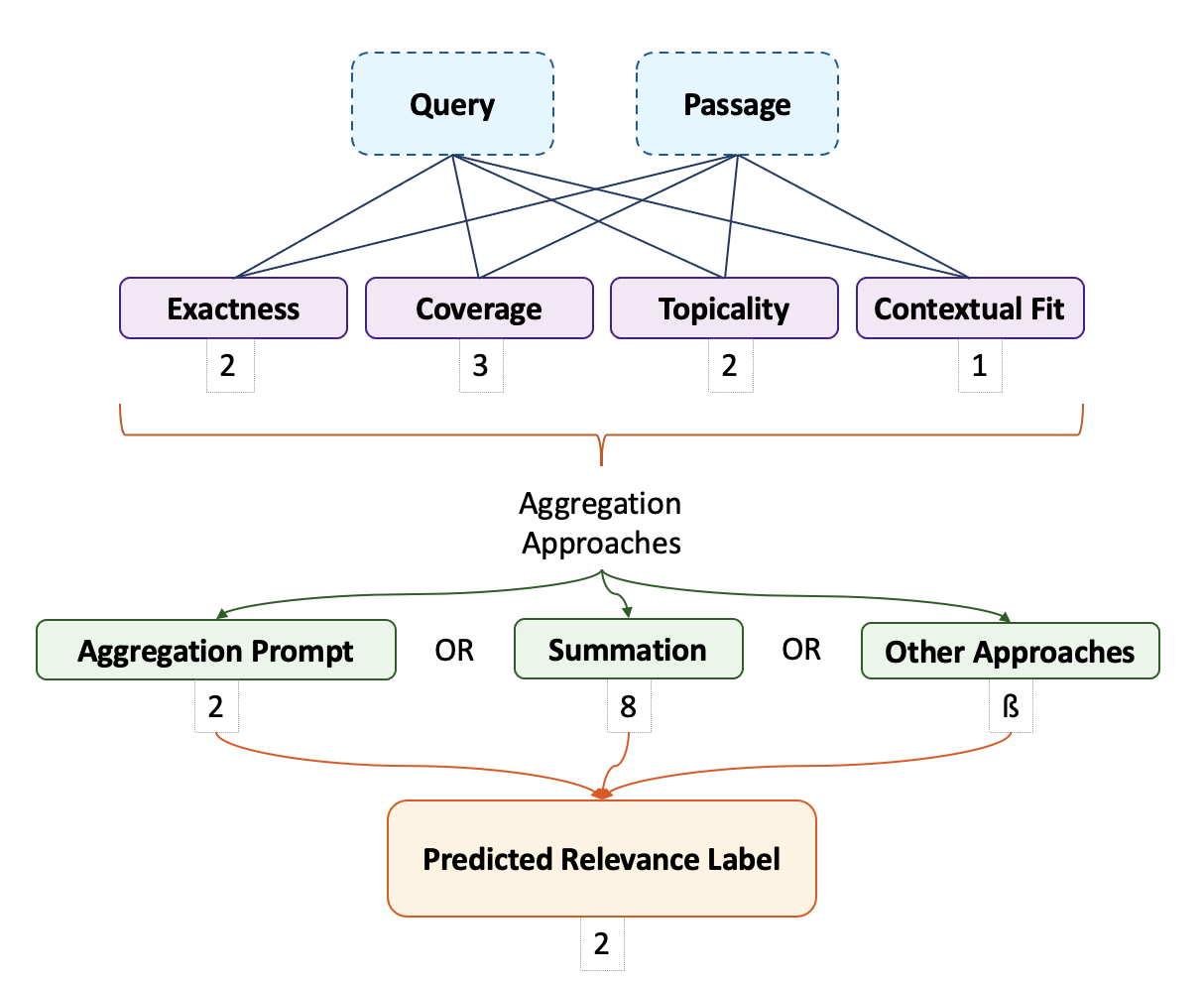}
    \caption{Overview the two phases of \fourprompts: grading criteria individually, then aggregating grades (see Section \ref{approach}).}
    \Description{Two-phase diagram showing criteria-specific grading followed by grade aggregation. Individual criteria like exactness are evaluated first, then combined using prompts or summation.}

    \label{fig:approach_overview}
\end{figure}

\paragraph{Issues with the current LLM Judge approaches.} Recent advances in large language models (LLMs) have shown promising potential for automated relevance assessment \cite{Sun2023IsCG, MacAvaney2023OneShotLF, Thomas2023LargeLM, Faggioli2023PerspectivesOL}. These approaches are based on the careful design of one LLM prompt, which for a given query-document pair can obtain a relevance label. Moreover, they have demonstrated strong empirical results in terms of leaderboard correlation, regularly obtaining Kendall's tau above 0.9, and an inter-annotator agreement Cohen's kappa of 0.40.

Grading with a single prompt, where the model is merely asked, ``Is this document relevant?'', comes with notable challenges. First, LLMs can sometimes produce unpredictable results, especially if the prompt design is not carefully controlled. Subtle variations in how a prompt is phrased may yield drastically different relevance predictions. What is worse is that this behavior changes across LLM families and versions. Second, the single-label output (``relevant'' or ``not relevant'') obscures the rationale behind the decision, limiting the transparency of the evaluation process. In IR and many other fields, interpretability is often vital: understanding why a decision was made can help diagnose systematic biases, guide system improvements, and build confidence among stakeholders. Finally, while LLMs may have been exposed to enormous corpora, their notions of relevance could deviate from established IR conventions or human judgments if they rely on obscure patterns in their training data.
Thus, while single-prompt approaches facilitate labeling large datasets, they are met with skepticism regarding their reliability, primarily due to a lack of transparency in the underlying reasoning.

\paragraph{Contribution.} To enable a more structured assessment, we propose a framework that evaluates relevance by decomposing it into multiple pre-defined criteria. Our approach assesses each criterion independently and then combines these evaluations to derive a final relevance label for the documents. While inspired by Chain-of-Thought (CoT) prompting \cite{Wei2022ChainOT}, our framework takes a distinctive approach by evaluating each criterion independently and in parallel, rather than following a sequential process. This approach allows us to investigate whether decomposing relevance into criteria can both improve evaluation quality and maintain competitiveness with existing LLM-based methods, while also making the grading process more transparent to human evaluators. We compare our method with recent frameworks, such as UMBRELA \cite{upadhyay2024umbrelaumbrelaopensourcereproduction}, which demonstrates better performance with larger LLMs \cite{farzi2025umbrela}. Our results show that, even with smaller LLMs, our approach provides equally effective evaluations.

\section{Related Work}

\subsection{LLM Based Relevance Judgment}
Recent work has explored using LLMs for automated relevance assessment. Many approaches rely on a single, hand-crafted prompt that the LLM uses to generate a relevance label.
Instances of this approach are as follows.
Thomas et al. \cite{Thomas2023LargeLM} demonstrate that LLMs can effectively generate relevance judgments through direct prompting, establishing a baseline for automated evaluation methods, like the UMBRELA framework \cite{upadhyay2024umbrelaumbrelaopensourcereproduction}.
Farzi and Dietz \cite{farzi2025umbrela} show that achieving optimal results with UMBRELA requires large LLMs, such as GPT-4.

Arabzadeh and Clarke \cite{arXiv:2504.12558} compare LLM-based methods (binary, graded, pairwise, and nugget-based) on TREC DL \cite{dl19, dl20} and ANTIQUE \cite{10.1007/978-3-030-45442-5_21} datasets, finding pairwise preferences best align with human labels while binary and graded (UMBRELA) methods achieve higher system ranking correlations. 

Sun et al. \cite{Sun2023IsCG} show zero-shot capabilities for relevance assessments. MacAvaney et al. \cite{MacAvaney2023OneShotLF} suggest one-shot learning for binary preference judgments. While Faggioli et al.\ \cite{Faggioli2023PerspectivesOL} show that LLM-based relevance assessment can be effective, they also warn of several pitfalls, including a potential mismatch between predicted labels and human-perceived relevance.

Beyond prompting, Zhang et al. \cite{zhang2024largelanguagemodelsgood} introduce a utility-based framework for QA datasets, outperforming relevance judgments. Meng et al. \cite{meng2024qppgenre} introduced QPP-GenRE based on fine-tuning LLMs to generate relevance judgments for query performance prediction on TREC DL datasets. However, Alaofi et al. \cite{Alaofi} show LLMs misclassify irrelevant passages due to lexical overlap, and Abbasiantaeb et al. \cite{abbasiantaeb2024uselargelanguagemodels} find that LLM-generated relevance judgments in conversational search tasks distort rankings when mixed with partial human annotations. These biases and inconsistencies show that single, coarse-grained metrics often fail to capture nuanced judgments.

\subsection{Grading Rubrics and Relevance Nuggets}
Recent advancements in LLM prompting, such as Wei et al.'s \cite{Wei2022ChainOT} chain-of-thought and Zhou et al.'s \cite{zhou_least--most_2023} least-to-most prompting, highlight the benefits of task decomposition. However, these techniques have not been used for the assessment of relevance criteria. Emerging approaches focus on decomposing relevance into smaller, LLM-verifiable points that are more interpretable to humans. The RUBRIC Autograder Workbench \cite{dietz2024workbench, farzi_pencils_2024} creates grading rubrics for each query, using answerability as a relevance indicator. The EXAM Answerability Metric \cite{DBLP:conf/desires/SanderD21} and EXAM++ \cite{farzi_exam_2024} uses multiple-choice questions and question-answering to test passage relevance. Other studies \cite{pradeep2024initialnuggetevaluationresults, mccreadie_crisisfacts_2023, pavlu_ir_2012} decompose queries into topical nuggets for relevance assessment. Our approach introduces a novel method for decomposing relevance based on specific criteria.

\subsection{Relevance Criteria in Classic IR}
    The conceptualization of relevance in information retrieval (IR) has evolved from a simple focus on topicality to a multi-criteria framework. Saracevic \cite{10.5555/1315930.1315947}, building on Cosijn and Ingwersen \cite{Cosijn/Ingwersen:00} and Borlund \cite{Borlund2003}, identified five dimensions of relevance: algorithmic, topical, cognitive, situational, and affective. Empirical studies, such as Boyce \cite{boyce_beyond_1982}, further support the view that relevance includes user needs, context, and emotional goals. Barry et al. \cite{barry_users_1998} and Xu et al. \cite{xu_relevance_2006} found that users rely on multiple criteria, like accuracy, specificity, and scope, when making relevance judgments, extending beyond topicality to include factors like novelty, reliability, and understandability. As conversational AI systems emerge, Sakai et al. \cite{sakai2023swangenericframeworkauditing} identified a schema of twenty-one evaluation criteria. However, previous frameworks were developed mainly for user-centric IR. Our work builds on these studies using criteria tailored for topical ad hoc retrieval.

\subsection{Research Gap}
Despite the progress in LLM-based relevance judgment and multi-criteria evaluation, a gap remains in integrating these approaches. Existing LLM evaluation methods often treat relevance as a monolithic concept, while traditional criteria-based approaches have not been adapted for automated assessment. Our work addresses this gap by proposing a framework that leverages LLMs' capabilities while preserving the interpretability and theoretical foundations of criteria-based evaluation.

\begin{table}
\centering
\caption{Chat completions prompt for criterion-specific grading. \textcolor{blue}{Variables}, denoted in blue with curly braces, are replaced with respective content.}
\label{tab:4prompts-phase1}
\begin{tabular}{p{0.45\textwidth}} 
\toprule
\textbf{Criterion-Specific Grading} \\ 
\midrule
\textbf{System Message:} \\ 
\small{
Please assess how well the provided passage meets specific criteria in relation to the query. Use the following scoring scale (0-3) for evaluation: 

0: Not relevant at all / No information provided. 

1: Marginally relevant / Partially addresses the criterion. 

2: Fairly relevant / Adequately addresses the criterion. 

3: Highly relevant / Fully satisfies the criterion. 
} \\
\midrule
\textbf{Prompt:} 
\small{
Please rate how well the given passage meets the \textcolor{blue}{\{Criterion Name\}} criterion in relation to the query. The output should be a single score (0-3) indicating \textcolor{blue}{\{Criterion Description\}}.

Query: \textcolor{blue}{\{Query\}}

Passage: \textcolor{blue}{\{Passage\}}

Score: 
} \\
\bottomrule
\end{tabular}
\end{table}

\begin{table}
\centering
\caption{Chat completions prompt used in Phase Two to aggregate criterion-level grades used into a final relevance label for each query-passage pair. \textcolor{blue}{Variables} are replaced by the corresponding grades obtained from Phase One.
\label{tab:4prompts-phase2}}
\begin{tabular}{p{0.45\textwidth}}
\toprule
\textbf{Aggregating Criterion Grades into a Relevance Label} \\
\midrule
\textbf{System Message:} \\ 
\small{
You are a search quality rater evaluating the relevance of passages. Given a 
query and passage, you must provide a score on an integer scale of 0 to 3
with the following meanings: 

3 = Perfectly relevant: The passage is dedicated to the query and contains
the exact answer. 

2 = Highly relevant: The passage has some answer for the query, but the
answer may be a bit unclear, or hidden amongst extraneous information. 

1 = Related: The passage seems related to the query but does not answer it. 

0 = Irrelevant: The passage has nothing to do with the query. 

Assume that you are writing an answer to the query. If the passage seems to
be related to the query but does not include any answer to the query, mark
it 1. If you would use any of the information contained in the passage in 
such an answer, mark it 2. If the passage is primarily about the query, or 
contains vital information about the topic, mark it 3. Otherwise, mark it 0.
}\\
\midrule
\textbf{Prompt:} \\
\small{
Query: \textcolor{blue}{\{Query\}}

Passage: \textcolor{blue}{\{Passage\}}

Exactness: \textcolor{blue}{\{Exactness Grade\}} 

Topicality: \textcolor{blue}{\{Topicality Grade\}} 

Coverage: \textcolor{blue}{\{Coverage Grade\}} 

Contextual Fit: \textcolor{blue}{\{Contextual Fit Grade\}}
\vspace{1ex} 

Please rate how the given passage is relevant to the query based on the
given scores. The output must be only a score (0-3) that indicates how relevant
they are. 

Score:
}\\
\bottomrule
\end{tabular}
\end{table}

\section{Approach}
\label{approach}
Our \fourprompts{} framework decomposes the notion of relevance into multiple fundamental criteria that together indicate whether a passage is relevant for a query.

In this work, we experiment with four criteria in particular: \exact{}, \cover{}, \topic{}, and \fit{}. This decomposition addresses two key challenges in LLM-based relevance judgments: the tendency of LLMs to conflate different aspects of relevance, and the need for interpretable judgments that are explainable and auditable by a human overseeing the evaluation process. Our approach consists of two phases that systematically separate individual criterion evaluation from the final relevance label aggregation. An overview of the approach is shown in Figure \ref{fig:approach_overview}.

\Description{Two-phase diagram showing criteria-specific grading followed by grade aggregation. Individual criteria like exactness are evaluated first, then combined using prompts or summation.}

\begin{description}
    \item[Phase One: Criterion Grading:] Each criterion is evaluated independently for every query-passage pair, enabling a focused assessment. 
    \item[Phase Two: Relevance Label Prediction:] After grading all criteria, we aggregate the individual criterion grades from Phase One to produce a relevance label that reflects the passage’s overall relevance for the query. This aggregation accounts for the semantic differences among different criteria, allowing for a more structured assessment of relevance.

\end{description}

A key advantage of our approach is that human evaluators can verify each criterion grade and its aggregation process, providing transparent oversight of automatic grading. We demonstrate this with an example in Section \ref{sec:qualitative}. Furthermore, the criteria and the aggregation method can be tailored to specific datasets and query types, offering flexibility across different application domains.

In the following we elaborate our \fourprompts{} approach in more detail.

\subsection{Phase One: Criterion-Specific Grading}
In the first phase, each query-passage pair is evaluated based on four relevance criteria elaborated below. These criteria align with diverse user expectations, reflecting various dimensions of relevance \cite{10.5555/1315930.1315947}. While these criteria are tailored towards TREC DL task, they can be adapted for different datasets and goals. We use the following criterion name and description as part of our prompt-based grading approach, as elaborated in Table \ref{tab:4prompts-phase1}.\footnote{Prompts are designed for chat completion models; for models that do not support chat completion, both messages can be concatenated.} Variables in the prompt templates (denoted in \textcolor{blue}{blue} and curly braces), such as criterion name or query, are replaced according to the criteria definitions. To ensure grammatical compatibility within the prompt format, the question mark in the Criterion Description is removed when used in the prompt.

\begin{description}
\item [Criterion Name: \textbf{Exactness}] $\,$\\
\textbf{Criterion Description}: How precisely does the passage answer the query?\\
\textbf{Rationale}: This criterion assesses whether the passage provides a specific, accurate response to the query’s core information need. It prioritizes targeted information over lexical overlap, ensuring relevance for goal-oriented or question-based queries.

\smallskip
\item [Criterion Name: \textbf{Topicality}]$\,$\\
\textbf{Criterion Description}: Is the passage about the same subject as the whole query (not only a single word of it)?\\
\textbf{Rationale}: This criterion ensures the passage addresses the entire query subject, aligning with topical relevance, which is determined by the degree of ``aboutness'' \cite{10.5555/1315930.1315947}, how closely the passage’s subject matches the topic expressed in the query. This guards against partial matches where only isolated terms from the query appear, rather than the full thematic content.

\smallskip
\item [Criterion Name: \textbf{Coverage}]$\,$\\
\textbf{Criterion Description}: How much of the passage is dedicated to discussing the query and its related topics?\\
\textbf{Rationale}: 

 A passage with high Coverage minimizes the user's need to seek additional information, aligning with the topical relevance through its focus on the breadth of relevant content. Additionally, it reflects a cognitive relevance \cite{10.5555/1315930.1315947} aspect: reducing the user's cognitive load by minimizing the need for further information seeking.

\smallskip
\item [Criterion Name: \textbf{Contextual Fit}]$\,$\\
\textbf{Criterion Description}: Does the passage provide relevant background or context?\\
\textbf{Rationale}: This criterion evaluates whether the passage offers relevant and appropriate background information that complements the overarching themes of the query. 

Contextual information helps resolve uncertainty and supports decision-making within a task-driven setting  \cite{10.5555/1315930.1315947}.

\end{description}

Each criterion is evaluated through dedicated prompts containing the criterion name and description, ensuring focused assessment of specific relevance aspects. Inspired by Chain-of-Thought prompting's principle of breaking down complex reasoning, this isolation strategy simplifies the task for LLMs by having them focus on one relevance aspect at a time rather than juggling multiple criteria simultaneously. This decomposition is particularly beneficial for smaller models, reducing task complexity and leading to more reliable criterion-specific evaluations. Additionally, the approach enhances transparency by creating clear links between individual prompts and their corresponding criterion assessments.

\subsection{Phase Two: Relevance Label Prediction}

Building on the criterion-based grades obtained in Phase One, this phase determines the overall relevance of the passage to the query by predicting a single relevance label. Various methods can be used for this aggregation, ranging from machine learning techniques to simple mathematical models. Here, we focus on a prompt-based aggregation approach that utilizes the criteria-specific grades.

\subsubsection{\fourprompts{}: Prompt-Based Aggregation}
Leveraging reasoning capabilities of LLMs, we employ an aggregator prompt, as outlined in Table \ref{tab:4prompts-phase2}, to instruct the LLM to evaluate the passage's overall relevance. This prompt integrates the query, passage, and criterion-specific grades, enabling the LLM to generate a comprehensive relevance label. We investigate the following combinations:

\begin{description}
\item[All four:] Using grades from all four criteria to produce a comprehensive relevance judgment.
\item[Three criteria:] Ablation study of three criteria to examine which subsets contribute most to effective relevance assessment.
\item[Single criterion:] Using each individual criterion as the relevance label to evaluate its standalone effectiveness.
\end{description}

Each criterion is evaluated with prompts of Table \ref{tab:4prompts-phase1} which include the query passage pair under evaluation, judgment instructions, and the exact criterion name and description. By standardizing prompt structures across all criteria while preserving their unique details, we enable meaningful comparisons across combinations.

\begin{table}
\centering
\caption{Mapping criterion grade sums to relevance labels used in Sumdecompose.
\label{tab:score_to_grade}}
\begin{small}

\begin{tabular}{lcccc}
\toprule 
\textbf{Criterion Grade Sum} & 10--12 & 7--9 & 5--6 & 0--4\tabularnewline
\midrule 
\textbf{Relevance Label} & 3 & 2 & 1 & 0\tabularnewline
\bottomrule
\end{tabular}
\end{small}
\end{table}

\subsubsection{Sumdecompose: Summation-based Aggregation:} An alternative is to sum the Phase One criterion grades and derive the final relevance label heuristically based on the total, using the mapping shown in Table~\ref{tab:score_to_grade}.

To assign a relevance label, we define grade thresholds over the total score, controlling the strictness of the evaluation. The thresholds used in our experiments are tuned on the \llmjudge{} development set (held out from evaluation) to maximize correlation with human judgments.

\FloatBarrier

\section{Experimental Evaluation}

\begin{table*}
\centering
\caption{Example from \llmjudge{} test set showing one query and two passages evaluated using \fourprompts{} grading with \llamasmall{}. For each passage, criterion grades (0-3) are assigned by \llamasmall{} in Phase-One via the Criterion-Specific prompts (see Table \ref{tab:4prompts-phase1}) and a predicted relevance label is obtained in Phase Two via the aggregation prompt (see Table \ref{tab:4prompts-phase2}). Manual relevance judgments are included from the dataset and for comparison. Rationales are written post-hoc, manually, and not generated by the LLM or drawn from the dataset.
The differences in \cover{} and \fit{} lead to different relevance predictions, illustrating nuances of the graded relevance labeling process.
\label{tab:examples_4prompts}}
\begin{small}
\begin{tabular}{p{0.14\linewidth} p{0.3\linewidth} p{0.14\linewidth} p{0.3\linewidth}}
\toprule
\multicolumn{4}{l}{\textbf{Query \texttt{3100119}} \hspace{1em} Do larger lobsters become tougher when cooked? }\\
\midrule
\multicolumn{2}{l}{\textbf{\texttt{msmarco\_passage\_01\_305447845}}} & \multicolumn{2}{l}{\textbf{\texttt{msmarco\_passage\_04\_612526438}}} \\
\midrule
\multicolumn{2}{p{0.48\linewidth}}{
\noindent\begin{minipage}[t]{0.5\textwidth}
\begin{quote}
Be sure to not overcook the lobster tails as the lobster meat will
be tough and chewy. A fully cooked lobster tail will have a bright
red shell and the lobster meat will be firm and white. \\
Serve with a squeeze of lemon, salt and pepper to taste.
\end{quote}\end{minipage}} 
& \multicolumn{2}{p{0.48\linewidth}}{
\noindent\begin{minipage}[t]{1\linewidth}
\begin{quote}
Saffitz takes it a step further, explaining that the best meat comes
from the lobster knuckle. The tail can be chewy and tough, she explains,
and the claws are stringy and often over-cooked. \\
\textquotedblleft The knuckle is already perfectly shredded, and
so tender,\textquotedblright{} she says.
\end{quote}\end{minipage}} \\
\midrule
\textbf{Exactness: 1} & No direct answer about size-toughness relationship, only mentions tail toughness from overcooking 
& \textbf{Exactness: 1} & No size-toughness relationship, only describes texture differences \\
\textbf{Coverage: 2} & Mostly focuses on toughness in cooking, partly on serving 
& \textbf{Coverage: 1} & Content split between chef attribution and comparing textures, with less focus on lobster cooking details \\
\textbf{Topicality: 2} & Mentions cooking and toughness, misses size aspect 
& \textbf{Topicality: 2} & Discusses parts and toughness, misses size aspect \\
\textbf{Contextual Fit: 2} & Provides visual indicators but lacks background on how cooking methods affect texture 
& \textbf{Contextual Fit: 1} & Lobster's part differences don't focus on cooking and/or size \\
\midrule
\multicolumn{2}{p{0.48\linewidth}}{
\textbf{Predicted Relevance Label: 2}  } & \multicolumn{2}{p{0.48\linewidth}}{\textbf{Predicted Relevance Label: 1 } } \\
\multicolumn{2}{p{0.48\linewidth}}{\textbf{Manual Judgment: 2}  }& \multicolumn{2}{p{0.48\linewidth}}{ \textbf{Manual Judgment: 1}} \\
\bottomrule
\end{tabular}
\end{small}
\end{table*}

We evaluate our \fourprompts{} approach to \llmevaluation{} with respect to the following research questions:

\begin{description}
    \item[RQ1:] How well does \fourprompts{} perform compared to other strong methods?
    \item[RQ2:] To what extent does the underlying LLM affect the results?
    \item[RQ3:] How effectively do the criteria
capture different aspects of relevance? 
    \item[RQ4:] How closely match predicted relevance labels the manual judgments?
    \item[RQ5:] What examples show that \fourprompts{} behaves as expected?
\end{description}

\subsection{Experimental Setup}
\subsubsection{Datasets}
We conduct our experiments on test sets of three major TREC datasets to ensure robust evaluation across different time periods and query types:
\begin{description}
    \item [\llmjudge{} (primary dataset)  \cite{rahmani_llmjudge_2024}:] 25 queries, 4,414 query-pas\-sage pairs, using systems submitted to the TREC Deep Learning Track of 2023. 
    We primarily focus our analysis on the \llmjudge{} dataset, as it represents the most recent public test collection and minimizes potential data leakage with LLM training.
    \item [\dltwenty{} \cite{dl20}:] 43 queries, 37 systems for passage retrieval from TREC Deep Learning track of 2020.
    \item [\dlnineteen{} \cite{dl19}:] 54 queries, 59 systems for passage retrieval from TREC Deep Learning track of 2019.
\end{description}

\subsubsection{Large Language Models.}
We evaluate our approach using three different openly available LLM models:

\begin{description}
    \item[\llamasmall{} (primary LLM):]\!\!\footnote{\url{https://huggingface.co/meta-llama/Meta-Llama-3-8B-Instruct}} A medium-scale public model that performs well with our approach.
    
    \item[\llamabig{}:]\!\!\footnote{\url{https://api.together.xyz/models/meta-llama/Llama-3.3-70B-Instruct-Turbo}} A recent large-scale public model from the same family.
    
    \item[\tfive{}:]\!\!\footnote{\url{https://huggingface.co/google/flan-t5-large}} A small-scale LLM used for fair comparison with the RUBRIC system~\cite{dietz2024workbench}.
\end{description}

Each model was used with identical prompting templates to ensure fair comparison. The temperature was set to 0.0 for deterministic outputs, and the maximum token length was set to 100 tokens. For model evaluation, we used \tfive{} and \llamasmall{} on an NVIDIA A40 GPU, and \llamabig{} through an API service \cite{togetherai2024}. 
To address RQ2 and RQ4, we compare our Multi-Criteria approach using these LLMs against a diverse set of LLMs and methods, including fine-tuned and prompt-based approaches, in the LLMJudge challenge (Table \ref{tab:main-results}).

\subsubsection{Evaluation Measures.} Our main objective is to assess if an \llmevaluation{} method can reliably identify which of many IR systems performs best. To accomplish this, we compare each system’s ranking under automatically generated relevance labels to its official ranking based on manual assessments. We measure how similar these two leaderboards are using rank correlation metrics:

\begin{description}
\item [Spearman’s rank correlation:] Reflects how well the relative ordering of systems aligns across the two leaderboards, taking into account each system’s exact rank position.
\item [Kendall’s tau:] Considers the consistency of pairwise system orderings between the two leaderboards.
\end{description}

\begin{table}
    % \centering{%
    \caption{Top 15 \llmjudge{} challenge results from Rahmani et al.~\cite{rahmani_llmjudge_2024}, evaluated using Spearman’s rank correlation and Kendall’s Tau, ordered by Spearman’s rank. Our proposed \fourprompts{} (denoted ``4prompts'') method, using \llamasmall{}, ranks first in Spearman’s correlation and second in Kendall’s Tau, outperforming many large-scale LLMs. Full results are available in Rahmani et al.~\cite{rahmani_llmjudge_2024}.
    \label{tab:main-results}
    }
\begin{small}

\begin{tabular}{lcc}
\toprule 
\textbf{Submission}  & \textbf{Spearman's rank}  & \textbf{Kendall's Tau}\tabularnewline
\midrule 
TREMA-4prompts \textbf{(ours) } & \textbf{0.9919}  & 0.9483 \tabularnewline
prophet-setting2  & 0.9914  & \textbf{0.9516} \tabularnewline
NISTRetrieval-instruct0  & 0.9907  & 0.9440 \tabularnewline
NISTRetrieval-instruct1  & 0.9907  & 0.9440 \tabularnewline
NISTRetrieval-instruct2  & 0.9907  & 0.9440 \tabularnewline
RMITIR-llama70B  & 0.9883  & 0.9353 \tabularnewline
TREMA-sumdecompose \textbf{(ours)}  & 0.9870  & 0.9300 \tabularnewline
Olz-multiprompt  & 0.9867  & 0.9267 \tabularnewline
llmjudge-thomas3  & 0.9867  & 0.9181 \tabularnewline
TREMA-all  & 0.9863  & 0.9138 \tabularnewline
llmjudge-simple1  & 0.9863  & 0.9181 \tabularnewline
TREMA-questions  & 0.9839  & 0.9095 \tabularnewline
TREMA-naiveBdecompose \textbf{(ours) }  & 0.9838  & 0.9128 \tabularnewline
Olz-halfbin  & 0.9830  & 0.9085 \tabularnewline
h2oloo-zeroshot1  & 0.9827  & 0.9181 \tabularnewline
... & ...& ...\tabularnewline
% prophet-setting1  & 0.9826  & 0.9042 \tabularnewline
% Olz-somebin  & 0.9822  & 0.9042 \tabularnewline
% h2oloo-fewself  & 0.9822  & 0.9085 \tabularnewline
% Olz-exp  & 0.9819  & 0.9009 \tabularnewline
% TREMA-direct  & 0.9819  & 0.9009 \tabularnewline
% NISTRetrieval-reason0  & 0.9810  & 0.9052 \tabularnewline
% NISTRetrieval-reason2  & 0.9810  & 0.9052 \tabularnewline
% llmjudge-simple3  & 0.9810  & 0.9052 \tabularnewline
% willia-umbrela1  & 0.9806  & 0.9009 \tabularnewline
% NISTRetrieval-reason1  & 0.9802  & 0.9009 \tabularnewline
% TREMA-CoT  & 0.9799  & 0.8956 \tabularnewline
% RMITIR-GPT4o  & 0.9798  & 0.8966 \tabularnewline
% willia-umbrela2  & 0.9769  & 0.887 \tabularnewline
% Olz-gpt4o  & 0.9758  & 0.8793 \tabularnewline
% RMITIR-llama38b  & 0.9758  & 0.8879 \tabularnewline
% llmjudge-thomas2  & 0.9750  & 0.8793 \tabularnewline
% willia-umbrela3  & 0.9730  & 0.8707 \tabularnewline
% TREMA-nuggets  & 0.9718  & 0.8664 \tabularnewline
% llmjudge-thomas1  & 0.9689  & 0.8664 \tabularnewline
% prophet-setting4  & 0.9608  & 0.8568 \tabularnewline
% h2oloo-zeroshot2  & 0.9604  & 0.8353 \tabularnewline
% TREMA-rubric0  & 0.9544  & 0.8276 \tabularnewline
% TREMA-other  & 0.9447  & 0.8276 \tabularnewline
\bottomrule
\end{tabular}

\end{small}

\end{table}

\begin{table*}
\centering
\caption{Evaluation metrics across datasets and our approaches compared to established baselines Thomas, Faggioli, and RUBRIC. Systems are evaluated with \texttt{trec\_eval} metrics NDCG@10, mean-average prediction (MAP), and mean reciprocal rank (MRR); leaderboard correlation measured in Spearman's rank (S) and Kendall's Tau (T). - marks unavailable results. }
\label{tab:evaluation_metrics}

\begin{footnotesize}

\begin{tabular}{lccccccccccccccccc}
\toprule
 &  & \multicolumn{4}{c}{\textbf{\llmjudge{}}} &  & \multicolumn{4}{c}{\textbf{dl2020}} &  & \multicolumn{4}{c}{\textbf{dl2019}} &  & Wins \tabularnewline
\midrule
 &  & \multicolumn{2}{c}{NDCG@10} & MAP  & MRR  &  & \multicolumn{2}{c}{NDCG@10} & MAP  & MRR  &  & \multicolumn{2}{c}{NDCG@10} & MAP  & MRR  &  & \tabularnewline
\cmidrule(lr){3-4}\cmidrule(lr){5-5}\cmidrule(lr){6-6}
\cmidrule(lr){8-9}\cmidrule(lr){10-10}\cmidrule(lr){11-11}
\cmidrule(lr){13-14}\cmidrule(lr){15-15}\cmidrule(lr){16-16}

\textbf{\llamasmall{}}&  & S & T & S  & S  &  &S & T & S  & S  &  & S & T & S  & S  &  & \tabularnewline
\midrule
\fourprompts{} (ours)       &        &\textbf{ 0.994} & \textbf{0.952} & 0.968 & 0.874 &  & 0.972 & 0.871 & 0.617 & 0.626 &  & \textbf{0.972} & 0.871 & 0.349 & 0.852 &  & 3 \tabularnewline

sumdecompose (ours)   &        & \textbf{0.990} & 0.936 & \textbf{0.972} &\textbf{ 0.960} &  & \textbf{0.983} & \textbf{0.905} & 0.879 & \textbf{0.910} &  & \textbf{0.972} & 0.865 & 0.671 & \textbf{0.943 }&  &  \textbf{7}\tabularnewline
\fourprompts{} (TCF)  (ours)     &        & \textbf{0.992}& \textbf{0.950} & \textbf{0.972} & 0.953 &  & 0.979	& 0.885 & 0.706 & 0.780 &  & \textbf{0.979} & \textbf{0.891} & 0.454 & 0.938 &  &  5\tabularnewline

Thomas \cite{Thomas2023LargeLM}               &        &   0.987	&0.915&0.964&0.833& &  0.973 & 0.870 &\textbf{0.917} &0.891&  &\textbf{ 0.971}	&0.879 &0.549 &0.898&  &2 \tabularnewline 
Faggioli \cite{Faggioli2023PerspectivesOL}             &        &    0.975	&0.892&0.963 &0.943 & &\textbf{0.982}&\textbf{0.909} &0.902& 0.866& & \textbf{0.974} & 0.874 &\textbf{0.750} &0.909&  &4\tabularnewline 
UMBRELA \cite{farzi2025umbrela} & & \textbf{0.989} & 0.931  & - & - &   & 0.973 & 0.870 & - & - & & \textbf{0.975} & \textbf{0.894} & - & - \tabularnewline

\tabularnewline
 
\textbf{\tfive{}}&  &  &  &   &   &  &  &  &   &   &  &  &  &   &   &  & \tabularnewline
\midrule
\fourprompts{} (ours)        &        & \textbf{0.988} &	\textbf{0.924} & \textbf{0.971} & \textbf{0.942} &  & 0.965	& \textbf{0.867} & 0.919 & 0.909 &  & \textbf{0.987}	&\textbf{0.923} &\textbf{0.880} & 0.844 &  & \textbf{8} \tabularnewline
sumdecompose (ours)   &        & 0.978	& 0.884 & 0.953 &0.923&  & 0.955	& 0.851 & 0.917 & 0.912 &  & \textbf{0.985}	& \textbf{0.920} & \textbf{0.881} & \textbf{0.889} &  & 4 \tabularnewline
Thomas  \cite{Thomas2023LargeLM}               &        &    \textbf{0.988}	& \textbf{0.924}          &   0.748    &     0.643  &&  0.936 &0.810 &0.828 &0.751&& 0.960 &0.833 &0.341 &0.755&  &2 \tabularnewline
Faggioli \cite{Faggioli2023PerspectivesOL}             &        &    0.984 &	0.912         &   0.965    &     0.938  &&  0.966 &0.861 &0.922 &\textbf{0.940}&& 0.968 &0.875 &0.864 &0.810&  &1\tabularnewline 
RUBRIC \cite{dietz2024workbench,farzi_pencils_2024}               &        &      \textbf{0.983}	&0.898     &   0.964    &    0.787   &  & \textbf{0.974} & \textbf{0.875} & \textbf{0.946} & 0.845 &  & 0.961 & 0.848 & \textbf{0.882} & 0.795 &  & 5 \tabularnewline
UMBRELA \cite{farzi2025umbrela} & & 0.971 & 0.868  & - & - &   & 0.880 & 0.717  & - & - & & 0.964 &0.862  & - & - \tabularnewline

\tabularnewline

\textbf{\llamabig{}}&  &  &  &   &   &  &  &  &   &   &  &  &  &   &   &  & \tabularnewline
\midrule
\fourprompts{} (ours)       &        & \textbf{0.994} & \textbf{0.952} & \textbf{0.968} & 0.911 & & 0.988 & 0.927 & 0.919 & 0.921 &  & 0.967 & 0.856 & 0.762 & 0.881 &   &  3\tabularnewline
sumdecompose  (ours)  &        & 0.988 &	0.939 & 0.963 &0.911&  & \textbf{0.992} &	\textbf{0.938} & \textbf{0.937 }& \textbf{0.965} &  &	0.965	& 0.856 & 0.814 & 0.915 &  & 4 \tabularnewline
\fourprompts{} (TCF) (ours)      &        & \textbf{0.994 }& \textbf{0.953} & \textbf{0.969} & 0.920 &  &\textbf{ 0.991}	& 0.928 & 0.922 & 0.910 &  & \textbf{0.975} & \textbf{0.879} & 0.770 & 0.896 &  & \textbf{6} \tabularnewline

Thomas \cite{Thomas2023LargeLM}               &        &    \textbf{0.991}	& 0.939  &   \textbf{0.970}   & 0.902 & &  0.922 & 0.929 &0.821 &0.887&  & 0.969&	0.863 &\textbf{0.873} &\textbf{0.925}&   &4\tabularnewline 

Faggioli \cite{Faggioli2023PerspectivesOL}             &        &    0.985&	0.912 & \textbf{0.968} & \textbf{ 0.958} &  & 0.980 &	0.901 &0.857 &\textbf{0.963}&  & 0.927&	0.767 &0.703 &0.857& &2 \tabularnewline 
UMBRELA \cite{farzi2025umbrela} & & \textbf{0.993} & 0.946  & - & - &   & 0.986 &0.911  & - & - & & 0.970 &0.869  & - & - \tabularnewline

\bottomrule
\end{tabular}

\end{footnotesize}

\end{table*}

\noindent Both measures range from -1.0 to +1.0, where higher correlation values indicate better alignment between the automatic relevance labels and the official human-labeled assessments.

\textbf{Significance analysis.} As rank correlation measures do not support a typical significance analysis in terms of standard-error or paired-t-tests, we consider small differences of 0.005 as equally good and highlight only differences of at least 0.025 to the best score.

Leaderboards are produced by using predicted relevance labels with \texttt{trec\_eval} using the following evaluation metrics
\begin{description}
    \item[NDCG@10 (primary measure): ] Normalized cumulative gain using a cutoff at rank 10. This is the primary measure  used in all TREC DL test collections.
    \item[MAP:] A mean of average-precision of all relevant passages.
    \item[MRR:] A mean of reciprocal rank of the first relevant passage.
\end{description}

While our approach is compatible with any number of desired relevance-criteria, here we study our \fourprompts{} approach with \exact{} \textbf{(E)}, \cover{} \textbf{(C)}, \topic{} \textbf{(T)}, and \fit{} \textbf{(F)} criteria, reflecting common aspects of relevance tailored to the TREC DL setting. Likewise, the aggregation step, whether
prompt-based, summation-based, or classifier-based is interchangeable, enabling the use of alternative strategies as needed.

\begin{figure}
    \centering
    \includegraphics[width=1\linewidth]{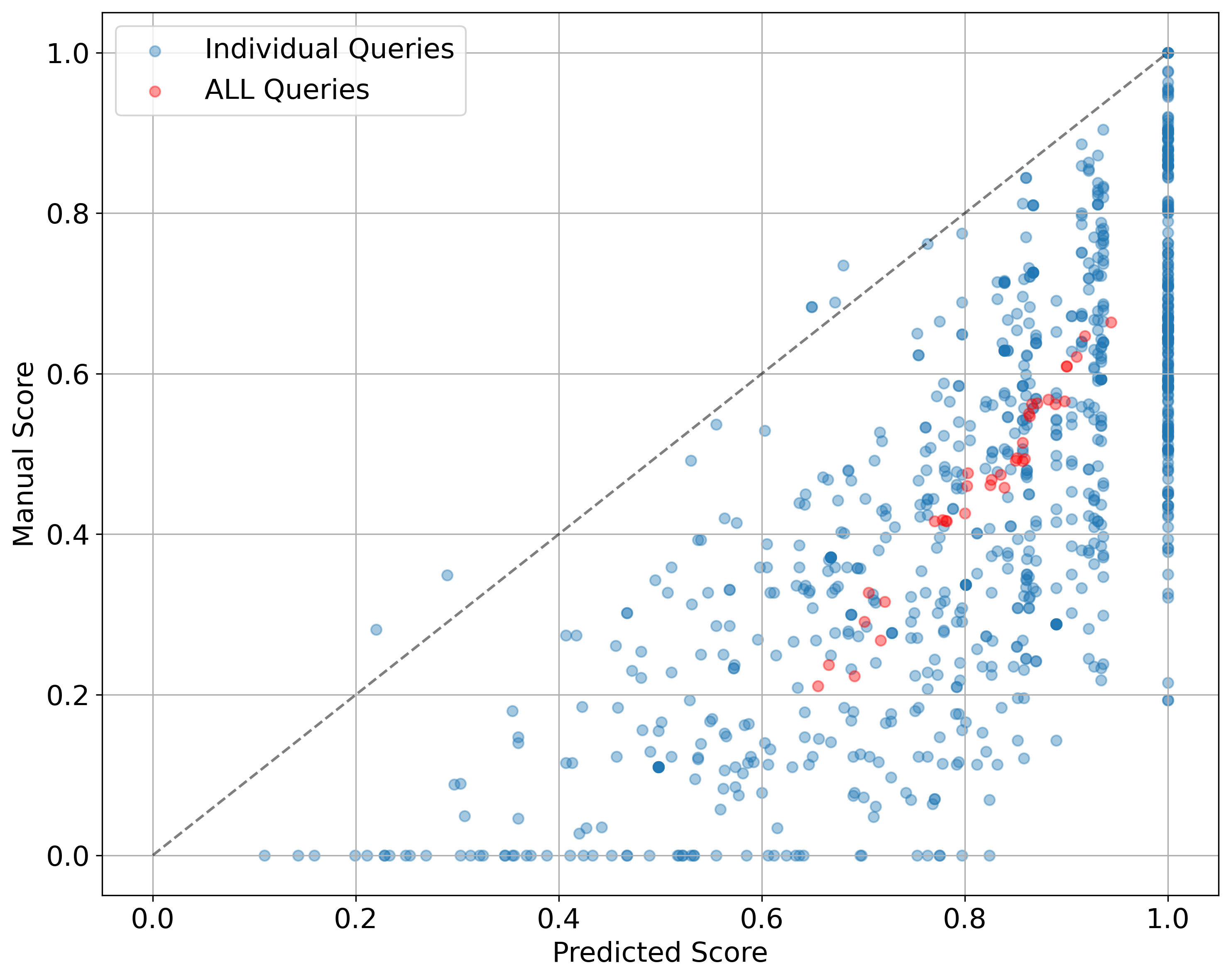}
    \caption{NDCG@10 evaluation scores under manual judgments vs.\ automatic relevance labels. Red dots represent the system performance, their ordering representing the leaderboard. For reference, all per-query/system evaluation scores are represented as blue dots. The red diagonal confirms that all systems are nearly in the same order under both automatic and manual methods, which results in a Spearman Rank coefficient of 0.99.
    \label{fig:scatter}
    }
    \Description{Scatter plot comparing NDCG@10 scores between manual judgments and automatic relevance labels, showing system performance (red dots) and per-query scores (blue dots)}
\end{figure}

\subsection{RQ1: Performance of \fourprompts{} Compared to Other Methods}

\paragraph{\llmjudge{} Challenge.} 
We evaluate our approach in the context of the First \llmjudge{} challenge (July 2024), comparing it against established \llmevaluation{} methods such as UMBRELA \cite{upadhyay2024umbrelaumbrelaopensourcereproduction} and RUBRIC \cite{dietz2024workbench}.
As shown in Table~\ref{tab:main-results}, our Multi-Criteria approach (here denoted ``4prompts'')  effectively reproduces the leaderboard derived from manual judgments. Accordingly, we focus on leaderboard correlation metrics Spearman's rank and Ken\-dall's tau.

Our \fourprompts{} approach, based on the smaller \llamasmall{} LLM, achieved the highest Spearman’s rank correlation and the second-highest Kendall’s tau in the \llmjudge{} Challenge, outperforming all other submissions~\cite{rahmani2025judgingjudgescollectionllmgenerated}. These included methods using larger LLMs (e.g., GPT-4, LLaMA-70B, T5), a variety of prompting and fine-tuning strategies, and complex ensemble techniques. This result highlights the effectiveness of our multi-criteria decomposition, demonstrating that breaking down complex information needs can yield strong performance—even with smaller models.

We also compare against UMBRELA~\cite{upadhyay2024umbrelaumbrelaopensourcereproduction} through a reproduction on \llamabig{}~\cite{farzi2025umbrela} (results in Table~\ref{tab:evaluation_metrics}), and find that \fourprompts{} performs at least as well. Notably, UMBRELA was originally designed for GPT-4o, which is a significantly larger model.

We also explore two variations: a sum-based aggregation of the four criteria grades (``sumdecompose'') and a Naive Bayes aggregation (``naiveBdecompose''). Despite relying solely on the smaller \llamasmall{} model and no additional fine-tuning, \fourprompts{} methods perform remarkably well compared to other top performing systems that utilize extensive fine-tuning (``setting2'') or larger LLMs (``llama70B,''   ``thomas3,'' ``umbrela2''). Recent work \cite{rahmani_judgeblender_2024}  shows that ensembles of large LLMs can achieve marginally higher performance.

\bigskip

\paragraph{Human vs.\ LLM label}
Figure \ref{fig:scatter} shows why our \fourprompts{} approach achieves such a high Spearman rank correlation. Each evaluated system’s performance is plotted as a red dot, with the horizontal axis showing scores derived from our predicted relevance labels, and the vertical axis showing scores from manual labels. The strictly ascending line of red dots indicates that both sets of labels yield nearly the same ranking of system quality, corresponding to a Spearman rank correlation of 0.99.

To provide a per-query perspective, each query-system score pair is shown as a blue dot. Notably, some query-system combinations receive zero points under manual assessments but a non-zero score under \fourprompts{}—visible as a horizontal line along the bottom of the figure. This pattern reflects the fact that \fourprompts{}, like many other \llmevaluation{} methods, tends to be more lenient than human judgments, a trend also observed in related work \cite{rahmani2025understandingbiassyntheticdata}.

\bigskip

\paragraph{\dltwenty{} and \dlnineteen{} datasets.}
Table~\ref{tab:evaluation_metrics} compares performance across multiple datasets, evaluation metrics, LLMs, and leaderboard correlation measures. We evaluate our methods against established prompt-based baselines, including Thomas~\cite{Thomas2023LargeLM}, Faggioli (B)~\cite{Faggioli2023PerspectivesOL}, RUBRIC~\cite{dietz2024workbench,farzi_pencils_2024}, and UMBRELA~\cite{upadhyay2024umbrelaumbrelaopensourcereproduction,farzi2025umbrela}.

To enable a fair comparison with UMBRELA, which was designed for the much larger GPT-4o, we use reproduction results~\cite{farzi2025umbrela} on the same \llama{} LLMs. While UMBRELA performs competitively on \dlnineteen{} with \llamasmall{}, it struggles to maintain consistent performance across datasets when applied to \tfive{}, where other methods show greater robustness.

Across 36 comparisons, the \fourprompts{} approach with prompt-based aggregation achieves the highest score more often than any other method---14 times---demonstrating particular strength on the \llmjudge{} dataset. On the simpler, earlier TREC Deep Learning (DL) collections, the ``sumdecompose'' variant is a strong contender.

In some cases, particularly on \llmjudge{} and \dlnineteen{}, the TCF variant, which uses only \topic{}, \cover{}, and \fit{} while skipping \exact{}, achieves better results. 
Faggioli’s yes or no prompt is notably effective under mean-reciprocal rank (MRR), whereas RUBRIC excels on the \dltwenty{} dataset and also performs well on \llmjudge{} when evaluating Spearman's rank with an NDCG@10 leaderboard.

\begin{table}
\centering
\caption{
Ablation study comparing different subsets of the four proposed criteria across three datasets and two LLMs, using prompt-based aggregation. Differences within 0.005 of the best score are considered non-significant (shown in \textbf{bold}), while scores that fall more than 0.025 below the best are marked in \textcolor{red}{red}. While some subsets occasionally outperform the full set, none show consistently better results. Overall, using all four criteria is the most reliable, falling behind in only 2 out of 12 experiments.
\label{tab:ablation}\label{tab:evaluation_metrics_43}\label{tab:evaluation_metrics_1}}
\begin{footnotesize}

\begin{tabular}{lccccccccc}
\toprule 
 &  & \multicolumn{2}{c}{\textbf{\llmjudge{}}} &  & \multicolumn{2}{c}{\textbf{dl2020}} &  & \multicolumn{2}{c}{\textbf{dl2019}}\tabularnewline
\midrule 
 &  & \multicolumn{2}{c}{NDCG@10} &  & \multicolumn{2}{c}{NDCG@10} &  & \multicolumn{2}{c}{NDCG@10}\tabularnewline

\multicolumn{2}{l}{\textbf{\llamasmall{}}}  &  S  & T  &  & S  & T  &  & S  & T \tabularnewline
\midrule 
All  four &  & \textbf{0.994 } & \textbf{0.952 } &  & \textit{0.972}  & \textcolor{red}{0.871}  &  & \textit{0.972 } & \textit{0.871 }\tabularnewline
\midrule 
TCF  \;\;(-E)&  & \textbf{0.992} & \textit{0.949 } &  & \textbf{0.979 } & \textcolor{red}{0.885}  &  & \textbf{0.979 } & \textbf{0.891 }\tabularnewline
ECF  \;\;(-T) &  & \textbf{0.992 } & \textit{0.942 } &  & \textbf{0.980 } & \textit{ 0.891 } &  & \textbf{0.976 } & \textit{0.883 }\tabularnewline
ETF  \;\;(-C) &  & \textbf{0.993 } & \textit{0.949 } &  & \textcolor{red}{0.954}  & \textcolor{red}{0.836}  &  & \textit{0.974 } & \textit{0.875 }\tabularnewline
ETC  \;\;(-F) &  & \textbf{0.993 } & \textit{0.945 } &  & \textit{0.967 } & \textcolor{red}{0.854}  &  & \textbf{0.977 } & \textbf{0.889 }\tabularnewline
\midrule
E &  & \textbf{0.989 } & \textit{0.932 } &  & \textbf{0.984 } & \textit{0.905 } &  & \textbf{0.980 } & \textbf{0.891 }\tabularnewline
T &  & \textbf{0.994 } & \textbf{0.955 } &  & \textit{0.967 } & \textcolor{red}{0.847}  &  & \textit{0.970 } & \textcolor{red}{0.863} \tabularnewline
C &  & \textcolor{red}{0.987}  & \textcolor{red}{0.926}  &  & \textbf{0.984 } & \textbf{0.909 } &  & \textit{0.962 } & \textcolor{red}{0.854} \tabularnewline
F &  & \textbf{ 0.990 } & \textit{0.935 } &  & \textbf{0.984 } & \textbf{0.911 } &  & \textbf{0.975 } & \textit{0.875 }\tabularnewline
\midrule 
\tabularnewline

\multicolumn{2}{l}{\textbf{\llamabig{}}}   &  &  &  &  &  &  & \tabularnewline

All  four &  & \textbf{0.994 } & \textit{0.952 } &  & \textit{0.988 } & \textit{0.927 } &  & \textit{0.967 } & \textcolor{red}{0.856} \tabularnewline
\midrule 
TCF   \;\;(-E)&  & \textbf{0.994 } & \textbf{0.953 } &  & \textbf{ 0.991 } & \textit{0.928 } &  & \textbf{0.975 } & \textit{0.879 }\tabularnewline
ECF  \;\;(-T) &  & \textbf{0.994 } & \textbf{0.956 } &  & \textit{0.986 } & \textcolor{red}{0.911}  &  & \textit{0.962 } & \textcolor{red}{0.844} \tabularnewline
ETF   \;\;(-C)&  & \textbf{0.995 } & \textbf{0.958 } &  & \textit{0.987 } & \textcolor{red}{0.917}  &  & \textit{0.958 } & \textcolor{red}{0.841} \tabularnewline
ETC   \;\;(-F)&  & \textbf{0.994 } & \textit{0.950 } &  & \textit{0.986 } & \textcolor{red}{0.911}  &  & \textit{0.971 } & \textcolor{red}{0.871} \tabularnewline
\midrule
E &  & \textit{0.987 } & \textcolor{red}{0.922}  &  & \textbf{ 0.995 } & \textbf{0.952 } &  & \textit{0.974} & \textit{0.883 }\tabularnewline
T &  & \textbf{0.993 } & \textit{ 0.950 } &  & \textit{0.972 } & \textcolor{red}{0.892}  &  & \textcolor{red}{0.947}  & \textcolor{red}{0.819} \tabularnewline
C &  & \textbf{0.992 } & \textit{0.946 } &  & \textbf{0.991 } & \textit{0.928 } &  & \textbf{0.980 } & \textbf{0.904 }\tabularnewline
F &  & \textbf{0.995 } & \textbf{0.955 } &  & \textit{0.987 } & \textcolor{red}{0.918}  &  & \textit{0.965 } & \textcolor{red}{0.856} \tabularnewline
\bottomrule
\end{tabular}

\end{footnotesize}
\end{table}

\subsection{RQ2: Impact of the Underlying LLM}

A key goal of \llmevaluation{} is to perform well even with smaller LLMs, such as \tfive{} and \llamasmall{}, which can be run on more affordable GPUs and require less inference time per query. As shown in Table \ref{tab:evaluation_metrics}, our \fourprompts{} approach consistently delivers strong performance across a range of model sizes and datasets. Notably, the gains of this approach are more pronounced for smaller models, suggesting that breaking relevance into multiple criteria provides the scaffolding needed to achieve high-quality judgments, typically attainable only by larger models under simpler prompts.

For instance, Thomas' prompt \cite{Thomas2023LargeLM} performs well with the larger \llamabig{} model but struggles with \llamasmall{}. In contrast, our \fourprompts{} approach not only secures first place in the \llmjudge{} challenge using smaller LLMs without fine-tuning, but also outperforms heavily fine-tuned methods (e.g., ``setting2'') and close\-ly approaches the performance of ensemble-based strategies such as LLMBlender \cite{rahmani_judgeblender_2024}. This highlights the effectiveness of multi-cri\-ter\-ion prompts in closing the performance gap between small and large language models.

Our focus is on achieving strong results with small LLMs, targeting scenarios where access to powerful GPUs or commercial API services is limited. Nevertheless, we compare the runtime of our \fourprompts{} approach with UMBRELA~\cite{upadhyay2024umbrelaumbrelaopensourcereproduction} in two settings: 5 queries $\times$ 5 passages and 10 queries $\times$ 10 passages.
Using \llamasmall{} in a chat-completion configuration, \fourprompts{} was $5.44-5.45\times$ slower than UMBRELA. In contrast, results for \tfive{} are comparable to UMBRELA, with only a smaller slowdown of $1.24-1.33\times$. 

The performance difference likely reflects several factors beyond prompt design, including model architecture, inference configuration, and processing overhead that accumulates across multiple query-passage pairs. Runtime could be further reduced through asynchronous parallelization, batching, and inference optimizations---potentially narrowing the gap while preserving output quality.

\begin{figure}
    \centering
    \includegraphics[width=1\linewidth]{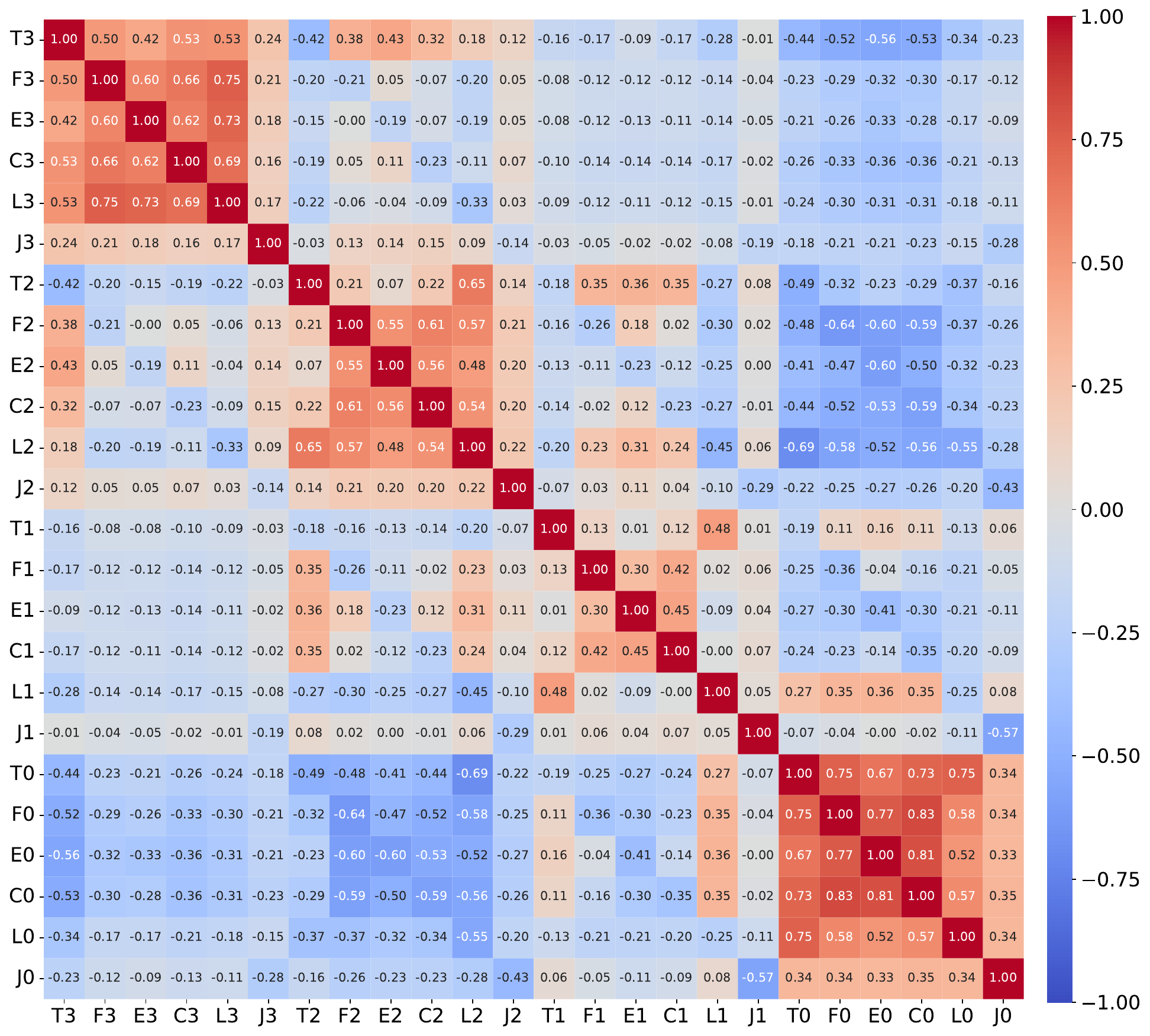}
    \caption{Abbreviations: T = Topicality, F = Contextual Fit, E = Exactness, C = Coverage, J = Ground Truth Relevance Judgment, L = Predicted Relevance Label. Numbers (0-3): indicate criteria-specific grades or relevance labels. Correlation matrix showing pairwise relationships between individual grade levels (0-3) across assessment criteria (Exactness, Coverage, Topicality, Contextual Fit) and their associations with predicted and ground truth relevance labels on \llmjudge{} dataset using \llamasmall{}.}
    \label{fig:correlation}
    \Description{Correlation matrix heatmap displaying relationships between assessment criteria (Topicality, Fit, Exactness, Coverage), their grade levels (0-3), and relevance labels for the LLM-Judge dataset using LLaMA-8B.}
\end{figure}

\subsection{RQ3: Different Aspects of Relevance}

To study whether all four criteria are necessary, we perform an ablation study in which one criterion is removed at a time, as well as experiments where only a single criterion is used. Table \ref{tab:ablation} shows these results, with top performers denoted in bold, while significantly worse methods marked in red.

We observe that, apart from two cases, the \fourprompts{} configuration with all four criteria either achieves or ties for the best performance, highlighting its robustness. Occasionally, subsets of three criteria—such as TCF on \dlnineteen{} and \dltwenty{}—slightly outperform the full set. Similarly, individual criteria can sometimes excel, for example \exact{} with \llamasmall{} or \cover{} with \llamabig{}. Nevertheless, no single criterion or subset is consistently better across all experiments.

\bigskip

To explore how different criteria inter-relate, Figure \ref{fig:correlation} presents the correlation matrix showing the relationships between \exact{} (E), \cover{} (C), \topic{} (T), and \fit{} (F),  as well as their alignment with predicted relevance labels (L) and manual judgments (J).

We observe a strong correlations across criteria, their predicted labels, and true judgments, particularly for non-relevant (0) and relevant labels (2 and 3).  Notably, a \topic{} score of 0 consistently signals non-relevance.

A few exceptions emerge for borderline relevance labels. For instance, label 1 often pairs with a \topic{} score of 1 but 0 grades in other criteria, while label 2 can coincide with 1 and 2 grades in some criteria and 2 or 3 in \topic{}. This pattern suggests that \topic{} can serve as a transitional marker between different levels of relevance. 

\Cover{} and \exact{} show strong positive correlations across all relevance levels, indicating that documents lacking exactness often fail to provide adequate coverage, yet they remain complementary for borderline relevance labels.

\Topic{} and \Fit{} also exhibit meaningful correlations, especially at level 0, suggesting that off-topic documents typically fail to fit the context as well.

\bigskip
Since the used Pearson correlation can hide differences in population sizes, we analyse the most frequent criteria-grade combinations. 
As expected, the all-zero pattern is most common, reflecting the large proportion of non-relevant labels.
The second most frequent pattern assigns a grade of 2 to most criteria, with a topicality score of 3, which aligns with earlier observations.

Overall, 29.6\% of cases assign only high criterion grades (greater than or equal to 2), 39.4\% assign only low grades (less than or equal to 1), and 31\% contain a mix of high and low values (for example, a \topic{} grade of 2 with a \fit{} grade of 1).

\begin{table}
\caption{ Inter-annotator agreement on \fourprompts{} with \llamasmall{}. Bold denotes highest counts per column; counts <10\% per row are in \lowcount{light gray}. 39\% of labels are correct, 38\% off-by-one, and Cohen's Kappa of 0 vs rest is 0.3. \label{tab:inter-ann-4prompt}}
\begin{small}
\begin{tabu}{@{}llccccrr@{}}%
\toprule%
&\textbf{Label}&\multicolumn{4}{c}{\textbf{Judgments}}&\textbf{Total}&\\%
\cmidrule(l@{\tabcolsep}){3-6}%
&&3&2&1&0&&\\%
\cmidrule(l@{\tabcolsep}){1-8}%

&3&\fbox{98}&97&116&122&433&\\%
&2&\textbf{243}&\fbox{\textbf{596}}&\textbf{682}&692&2213&\\%
&1&\lowcount{26}&\lowcount{72}&\fbox{244}&409&751&\\%
&0&\lowcount{10}&\lowcount{43}&191&\fbox{\textbf{771}}&1015&\\\bottomrule%
\end{tabu}%
\end{small}

\end{table}

\subsection{RQ4: Agreement With Human Judgments}

Figure \ref{fig:correlation} shows that predicted relevance labels have moderate to strong correlations with manual judgments for labels 0, 2, and 3 (approximately 0.34, 0.22, and 0.17, respectively), but are weaker for label 1. 

Table \ref{tab:inter-ann-4prompt} further indicates that our \fourprompts{} approach is generally more lenient than human annotators: when \fourprompts{} predicts a relevance label of 0, manual judgments usually concur, and when human judgments are high, \fourprompts{} also predicts high labels as well. Among 4,412 query-document pairs, there are 1,009 instances where the predicted label differs by at least two levels from the human judgment; in 92.2\% of these cases, \fourprompts{} is too lenient. A closer look at zero-labeled documents predicted as label 2 suggests these are borderline on-topic responses that omit important query details (see Section \ref{sec:qualitative} for examples). With additional fine-tuning or specialized classification, such mispredictions may be further reduced.

\subsection{RQ5: Illustrative Examples}
\label{sec:qualitative}

To illustrate the \fourprompts{} approach, Table~\ref{tab:examples_4prompts} presents two passages retrieved for a query about lobster preparation, along with their criterion-specific scores (on a 0--3 scale), predicted relevance labels, and manual judgments.
According to the output of the \fourprompts{} approach, both passages are broadly on topic (\topic{} = 2), but one shows better Coverage by including more cooking-specific content and demonstrates improved Contextual Fit. This subtle difference leads to a higher predicted relevance label, matching the manual scores of 2 and 1, respectively.
To clarify the reasoning behind each score, post-hoc rationales were written manually (not generated by an LLM and not part of the prompt output) for each criterion. These rationales highlight how differences in aspects like Coverage can influence overall relevance. This example shows how decomposing relevance into distinct criteria---here, \topic{}, \exact{}, \cover{}, and \fit{}---supports a more transparent and human-aligned labeling process. It also facilitates iterative refinement and more nuanced system evaluation.

\section{Conclusion}
We present the \fourprompts{} approach to \llmevaluation{}, which decomposes the notion of relevance into multiple criteria that are individually graded and subsequently aggregated into a single relevance label. In our experiments, we use the criteria \exact{}, \cover{}, \topic{}, and \fit{}. This approach achieves the highest Spearman's rank correlation (0.99) in the first \llmjudge{} challenge of July 2024, outperforming strong contenders such as UMBRELA and RUBRIC.

In our experiments spanning three datasets, three LLMs, and several reference baselines, \fourprompts{} consistently delivers the strongest results. When examining the correlations among individual criteria, predicted labels, and judgments, we find that \topic{} is essential but not by itself sufficient. Rather, the final relevance label gains robustness from the complementary nature of the different criteria. Although \fourprompts{} tends to assign higher relevance scores than human annotators, this leniency applies uniformly across all tested systems. For example, when \fourprompts{} is used to rank IR systems by quality, it produces a leaderboard that closely matches the one based on manual judgments. This is evident from the line of red dots in Figure~\ref{fig:scatter}, corresponding to a Spearman's rank correlation of 0.99.

One notable advantage of \fourprompts{} is that it not only remains computationally accessible—even with smaller LLMs such as \llamasmall{} and \tfive{}—but also delivers consistently strong results. In contrast, many top contenders in the \llmjudge{} challenge depend on significantly larger models and extensive fine-tuning. By offering an \llmevaluation{} approach that performs exceptionally well without requiring expensive hardware, we aim to broaden research access for under-resourced communities and enhance the reproducibility of IR research.

\section{Limitations}

As with all LLM-based evaluators, our empirical meta-evaluation has several limitations, which we outline using the framework of LLM Evaluation Tropes~\cite{dietz2025tropes}.

First, our meta-evaluation is based on IR systems submitted to the TREC Deep Learning tracks prior to 2024 (\textit{Old System Trope}). The next generation of systems are likely to incorporate LLM-based components within retrieval or generation, potentially resulting in different empirical behavior and introduces a risk of circularity.

Second, because the training data for large language models is not publicly available, it is possible that the LLMs used in our study were exposed to content from the TREC DL test collections. This raises concerns about evaluation signal leakage, as described by the \textit{Test Set Leak Trope}, which could inflate performance estimates.

Finally, LLMs are frequently retrained or updated, which may cause their relevance judgments to shift over time and diverge from those of human annotators. This dynamic is captured by the \textit{LLM Evolution Trope}. We therefore recommend continued meta-evaluation of LLM-based assessors, especially when applied to new domains, model versions, or IR architectures.

\begin{acks}
We thank the organizers of the \llmjudge{} challenge \cite{rahmani_llmjudge_2024} for the opportunity to participate and their support in sharing the data  necessary to produce Table \ref{tab:main-results}.

    This material is based in part upon work supported by the National Science Foundation under Grant No. 1846017. Any opinions, findings, and conclusions or recommendations expressed in this material are those of the authors and do not necessarily reflect the views of the National Science Foundation.
\end{acks}

\FloatBarrier

\bibliographystyle{plain}
\bibliography{sample}
\end{document}